\documentclass[aps,prl,reprint,longbibliography]{revtex4-1}
\usepackage{graphicx}
\usepackage[colorlinks, linkcolor=blue, urlcolor=blue, anchorcolor=blue, citecolor=blue]{hyperref}

\begin{document}

\title{Plasmonic Purcell Effect in Organic Molecules}
\author{D. Zhao$^{1,2}$}
\author{R. E. F. Silva$^2$}
\author{C. Climent$^2$}
\author{J. Feist$^2$}
\author{A. I. Fern\'{a}ndez-Dom\'{\i}nguez$^2$}
\author{F. J. Garc\'{\i}a-Vidal$^{2,3}$}
\affiliation{
$^1$School of Physical Science and Technology, Southwest University, Chongqing 400715, China \\
$^2$Departamento de F\'{\i}sica Te\'{o}rica de la Materia Condensada and Condensed Matter Physics Center (IFIMAC), Universidad Aut\'{o}noma de Madrid, E-28049 Madrid, Spain\\
$^3$Donostia International Physics Center (DIPC), E-20018 Donostia/San Sebasti\'{a}n, Spain}

\date{\today}
\begin{abstract}
By means of quantum tensor network calculations, we investigate
the large Purcell effect experienced by an organic molecule placed
in the vicinity of a plasmonic nanostructure. In particular, we
consider a donor-$\pi$ bridge-acceptor dye at the gap of two Ag
nanospheres. Our theoretical approach allows for a realistic
description of the continua of both molecular vibrations and
optical nanocavity modes. We analyze both the exciton dynamics and the corresponding emission spectrum, showing that 
these magnitudes are not accurately represented by the simplified models used up to date. By disentangling
the molecule coupling to radiative and non-radiative plasmonic
modes, we also shed light into the quenching phenomenology taking
place in the system.
\end{abstract}

\maketitle

The Purcell effect~\cite{Purcell1946} lies at the core of quantum
electrodynamics, as it reveals that the radiative properties of
any quantum emitter are not inherent to it, but also depend on the
electromagnetic (EM) vacuum in its surroundings. A few decades
ago, this phenomenon, and in particular, the pursuit for the full
inhibition of spontaneous emission at EM band gaps, was one of the
driving forces behind the development of photonic
crystals~\cite{Yablonovich1987}. More recently, much interest has
focused on metallic nanocavities~\cite{Pelton2015,Mikkelsen2014}.
The large and spectrally complex photonic density of states
associated to plasmonic resonances allows an unprecedented control
over spontaneous emission in these
nanostructures~\cite{vanHulst2011,Giannini2011}. In recent years,
different light sources have been used to probe plasmonic Purcell
enhancement phenomena, such as quantum
dots~\cite{Belacel2013,Rakovic2015}, solid-state color
centers~\cite{Andersen2017,Bogdanov2018} or passing electron
beams~\cite{Kaminer2017,Martin-Jimenez2020}.

Due to their large transition dipole moments and stability at room
temperature, organic molecules have received increasing attention
as reliable quantum emitters. In their interaction with plasmonic
nanostructures, several topics have been addressed, such as
fluorescence
enhancement~\cite{Novotny2006,Kuhn2006,Muskens2007,Moerner2009,Acuna2012},
spectral shaping~\cite{LeRu2007, Ringler2008, Enderlein2009, Dong2010, Rivas2018,
Sandoghdar2019, Menon2020} and strong coupling~\cite{Bellessa2004,
Shi2014, Sanvitto2016,Baumberg2016, Wang2017, Baumberg2019}.
Unlike other microscopic light sources, for which a two-level
system (TLS) description is usually accurate, electronic
transitions of organic molecules, from now on {\it excitons},
interact strongly with the molecular nuclear vibrations. 

Within the TLS approach, only a few timescales play an important role, and in
particular, the relative values of the plasmon-exciton exchange rate and 
their respective losses can be used to distinguish between the weak-coupling and strong-coupling
regimes in a fundamental analysis. In contrast, in molecules, the vibronic
coupling and the vibrational dynamics introduce several new timescales, such as
the coherent nuclear oscillation period or the rate of vibrational energy
dissipation and thermalization time, and the simple dichotomy between weak and
strong coupling can be expected to give way to a richer phenomenology. For the
case of not too large Purcell factors, radiative decay is typically slower than
the vibrational relaxation and thermalization times (on the order of $1$~ps or
less~\cite{May2011}), and molecular emission can be assumed to proceed from the
lowest excited molecular state, with all vibrational modes in thermal
equilibrium. This assumption gives rise to a Fermi's golden rule (FGR)-based
approach in which the emission spectrum of the organic molecule near a plasmonic
structure is given by the product of its free-space spectrum and a
frequency-dependent Purcell
enhancement~\cite{Enderlein2002,Ringler2008,Enderlein2009}. However, as
plasmonic structures can achieve Purcell factors on the order of $10^6$ \cite{Baumberg2016},
radiative decay rates can be decreased from their free-space values in the
nanosecond range to femtoseconds, such that radiative decay
does \emph{not} proceed from vibrationally relaxed molecules, the FGR
approximation breaks down, and non-equilibrium effects play an important role.
More recently, cavity-QED approaches have been able to incorporate the interplay
between the exciton and one (or a few) vibrational
modes~\cite{Genes2019,Nazir2019,Kalle2019, Wang2019,Liang2020}. However, a realistic description of typical organic molecules 
could require considering hundreds of vibrational modes.

In this Letter we present an accurate theoretical framework that
is able to treat on an equal footing the electronic, vibrational and
plasmonic degrees of freedom associated with the Purcell effect
experienced by organic molecules placed within a metallic
nanocavity. Our numerical scheme is based on a quantum tensor
network (TN) method~\cite{Schroder2016,DelPino2018}, which helps
provide a complete picture of this phenomenon. In addition, it
allows for a validation of the simple models introduced above and
also for an in-depth study of the influence of the large number of
vibrational modes on the spontaneous emission rate and emission
spectrum. We show that while the TLS model is able to reproduce
the very short timescales of the spontaneous decay, the FGR approach
accounts only for the behavior at very long times. On
intermediate time scales, a model that incorporates a single
vibrational mode is able to extent the validity of the TLS to
slightly longer times. However, we demonstrate that it is
mandatory to account for the whole set of vibrational modes to
capture most of the time dynamics and the corresponding spectrum.

\begin{figure}
\includegraphics[width=\columnwidth]{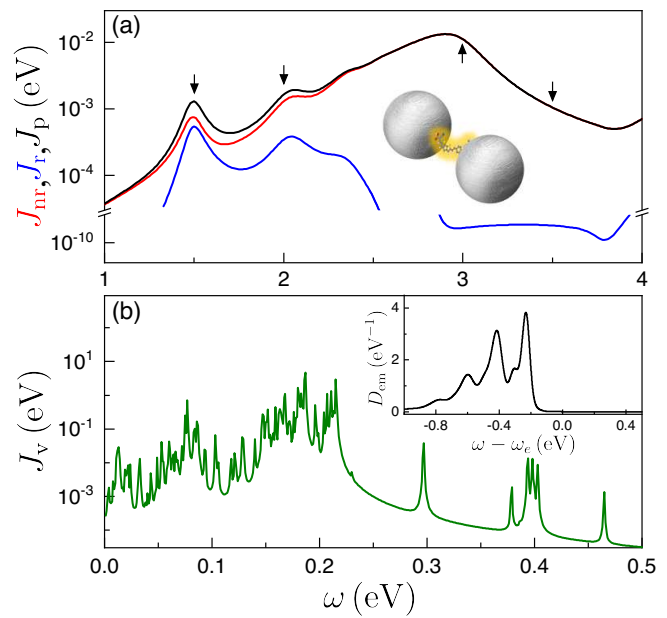}
\caption{\label{Schematic} (a) Radiative ($i = \textrm{r}$, blue) and non-radiative ($i = \textrm{nr}$, red) plasmonic spectral densities, evaluated at the gap center of a nanosphere dimer, sketched as an inset. The black curve shows the total spectral density, $J_{\mathrm{p}}(\omega)$, while the arrows indicate the exciton frequencies in Figs.~\ref{Dynamic} and~\ref{Spectra}. (b) Vibrational spectral density, $J_{\mathrm{v}}(\omega)$, for the CPDT molecule. The inset shows the lineshape function, $D_{\mathrm{em}}(\omega)$, at $T=300$ K.}
\end{figure}

As a case study, we consider the excited-state dynamics of a single organic molecule placed at the gap center of a silver nanosphere dimer, as sketched in the inset of \autoref{Schematic}(a). This plasmonic structure resembles  the two geometries in which the largest Purcell factors have been reported, the so-called nanoparticle on mirror~\cite{Mikkelsen2014} and bowtie antenna~\cite{Moerner2009} geometries. In this work we have chosen a donor-$\pi$ bridge-acceptor (D-$\pi$-A) organic dye, labeled CPDT \cite{Climent2013}, as a prototypical organic molecule because it displays a significant transition dipole moment and a large Stokes shift $(>$ 0.4 eV), yielding excellent absorption and emission capabilities. The Hamiltonian for this hybrid system can be written as [$\hbar=1$]
\begin{eqnarray}
\label{Hamiltonian}
H & & =  \omega_{e}\sigma_{+}\sigma_{-} \nonumber + \int d\omega \left [\omega b^{\dagger}_{\omega} b^{}_\omega + \lambda_{\omega}\left (b^{\dagger}_{\omega} +  b^{}_{\omega}\right )\sigma_{+}\sigma_{-} \right] + \nonumber\\
& & \sum_{i={\mathrm{nr, r}}} \int d\omega \left[\omega a_{i, \omega}^{\dagger}a_{i, \omega}^{} + g_{i, \omega} \left (a_{i, \omega}^{\dagger}\sigma_{-}+a_{i, \omega}^{}\sigma_{+}\right) \right],
\end{eqnarray}
where $\sigma_{-}$, $b^{}_{\omega}$ and $a_{i, \omega}^{}$ denote the lowering operator of the molecular electronic transition, annihilation operator for the molecular vibration and $i$-component of the plasmonic mode at frequency $\omega$. Here the index $i$ is either $\mathrm{nr}$ or $\mathrm{r}$, representing the non-radiative and radiative plasmonic modes, respectively. Notice that only the latter can be detected in the far-field.
The first line in Eq. (\ref{Hamiltonian}) describes the free-standing molecule, where we have used a Holstein-type Hamiltonian~\cite{Holstein1959} accounting for the continuum of vibrational modes, with coupling strengths $\lambda_{\omega}$. The exciton frequency $\omega_{e}$ corresponds to the vertical transition from the vibrational ground state of the electronic ground state. The second part in Eq. (\ref{Hamiltonian}) describes the continuum of radiative and non-radiative plasmonic modes, and their coupling, weighted by $g_{i, \omega}$, to the molecular exciton. This plasmon-exciton coupling can be encoded in the spectral density $J_{i}(\omega) = g_{i, \omega}^{2}$, in a similar way as the vibrational spectral density $J_{\mathrm{v}}(\omega) = \lambda^{2}_{\omega}$ does for the vibronic coupling.

We numerically solve both the excited-state dynamics and the emission spectrum with our quantum TN method. To apply this TN approach, the two continua in Eq. (\ref{Hamiltonian}) are transformed to a chain form in which each continuum (EM and vibrational) is represented by a chain of nearest-neighbor coupled oscillators, with only the first oscillator coupled to the exciton. More details of our TN numerical framework can be found in the Supplemental Material \cite{SM}. The first site in the vibrational chain is usually termed the reaction coordinate (RC), whose frequency is given by
\begin{equation}
\omega_{\mathrm{RC}} = \frac{1}{\lambda_{\mathrm{RC}}^{2}}\int \omega J_{\mathrm{v}}(\omega) d\omega,
\end{equation}
with $\lambda_{\mathrm{RC}} = \sqrt{\int J_{\mathrm{v}}(\omega) d\omega}$ being the coupling strength between the RC and the exciton. If the molecule is initially in its excited state, the coupling to the plasmonic chain leads to its decay whereas the coupling to the vibrational chain {\it dresses} the electronic state. 

To shed light into the effect of molecular vibrations and uncover the relevant timescales in the spontaneous emission process, we compare our TN numerical results against three simplified models. As the simplest choice, by discarding all the molecular vibrations, the standard TLS model predicts a decay only dictated by the plasmonic environment, $\gamma_{\mathrm{TLS}}=2\pi J_{\mathrm{p}}(\omega_e)$, where $J_{\mathrm{p}}(\omega) = \sum_{i} J_{i}(\omega)$ is the total plasmonic spectral density. In a second step, by keeping only the RC within the vibrational chain, we can derive an approach, dubbed here as the single vibration mode (SVM) approximation, in which the RC comprises all the vibrational response. As a difference, within the FGR model \cite{SM}, the spontaneous emission rate is simply calculated as the spectral integral of the product of $J_{\mathrm{p}}(\omega)$ with the so-called lineshape function, $D_{\mathrm{em}}(\omega)$, which represents the available optical transitions connecting the ground and the excited electronic states of the molecule under the assumption that the vibrational modes are in thermal equilibrium, leading to
\begin{equation}
\gamma_{\mathrm{FGR}} = 2 \pi \int d\omega D_{\mathrm{em}}(\omega) J_{\mathrm{p}}(\omega).
\end{equation}

As commented above, we consider a silver nanosphere dimer as an example of plasmonic cavity, with a $1$ nm gap and $20$ nm radius, embedded in a matrix with refractive index $n_{\mathrm{D}} = 2.1$. The dielectric constant for silver is taken from Ref.~\cite{Palik1998}. As shown in \autoref{Schematic}(a), the plasmonic spectral density is characterized by a radiative dipole mode located at around $1.5$ eV, a quadrupole mode at $2.1$ eV, and a dominant non-radiative mode, the so-called pseudomode~\cite{Antonio2016}, emerging at $2.9$ eV. A single CPDT molecule is located at the gap center and we assume that its transition dipole moment (modulus $\mu = 0.1$ $e\cdot$nm) is pointing along the line that connects the two nanospheres. In the numerical calculations, we assume that the initial state originates from a Franck-Condon excitation, i.e., a vertical transition that could result after an ultrashort laser pulse excitation of the electronic ground state.

Electronic structure calculations within density functional theory and its time-dependent version are performed to obtain the vibronic coupling constants of the CPDT dye for the displaced harmonic oscillator model. Both the vibrational spectral density and lineshape function of the CPDT molecule used in our calculations are shown in \autoref{Schematic}(b). In order to explore different regions of the plasmonic spectral density while keeping the vibrational structure unchanged, we artificially shift the (vertical) exciton energy of the molecule to different values.

\begin{figure}
\includegraphics[width=\columnwidth]{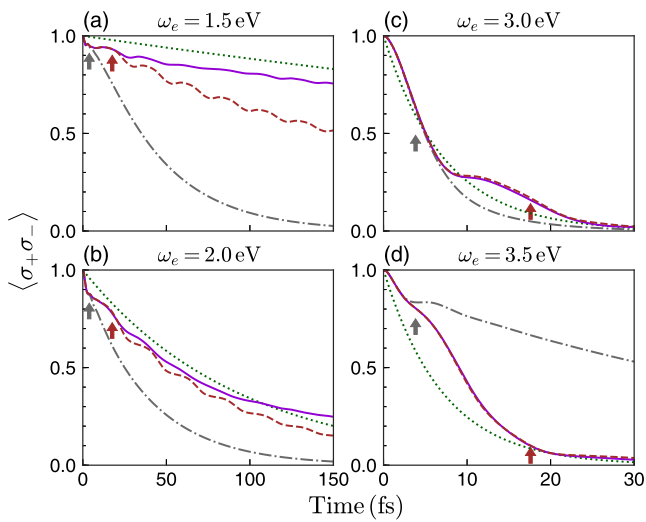}
\caption{\label{Dynamic} Excited-state dynamics $\langle \sigma_{+} \sigma_{-} \rangle$ versus time for different exciton frequencies. In each panel, TLS  (gray dash-dotted), SVM  (brown dashed), FGR (green dotted), and exact TN (violet solid) predictions are shown. The arrows indicate the positions of $\tau_{\mathrm{1}}$ (gray) and $\tau_{\mathrm{2}}$ (brown).}
\end{figure}

We focus first on the excited-state dynamics. Figure~\ref{Dynamic} shows the evolution of the exciton population, $\langle \sigma_{+} \sigma_{-} \rangle$, evaluated with the TN method at four exciton frequencies. It is compared against the results of the three different models (TLS, SVM and FGR) for the same cases. First, it is evident that the TLS model largely fails to describe the decay dynamics for all the cases depicted in \autoref{Dynamic}. This failure highlights the importance of going beyond the TLS approach when defining the coupling regimes in the interaction of organic molecules with nanophotonic structures, implying that the simple distinction between weak and strong coupling has to be modified for organic molecules. Still, it is interesting to note that TLS seems to be valid up to times of about $\tau_{\mathrm{1}}\approx4$~fs (grey arrows in \autoref{Dynamic}). This is because, soon after its generation, the exciton vibrational wavepacket remains in the vicinity of the Franck-Condon region before exploring the potential energy surface of the excited state~\cite{Silva2019}. Hence, within a very short timescale, vibrations do not play any role yet and the TLS model describes the molecular decay. This time scale, $\tau_{\mathrm{1}}$, can be estimated as a fraction of the period associated with the RC harmonic motion, $T_{\mathrm{RC}}=2\pi/\omega_{\mathrm{RC}}$, with $\tau_{\mathrm{1}} \approx T_{\mathrm{RC}}/6 = 3.8$~fs giving a good estimate of this initial timescale.

When the wavepacket initially moves away from the Franck-Condon region, its dynamics is dominated by the RC, as revealed by the accuracy of the SVM description for all $\omega_{e}$ in \autoref{Dynamic}. This regime holds until both dephasing and decay of the RC into other sites in the TN vibrational chain becomes important. We can estimate this second timescale, $\tau_{\mathrm{2}}$, as the inverse of the coupling between the first (RC) and second sites of the vibrational chain,
\begin{equation}
\gamma_{\mathrm{d}} = \sqrt{\frac{1}{\lambda_{\mathrm{RC}}^{2}}\int \omega^{2}J_{\mathrm{v}}(\omega) d\omega - \omega_{\mathrm{RC}}^{2}}.
\end{equation}
This estimation gives $\tau_{\mathrm{2}}=1/\gamma_{\mathrm{d}}=17.6$ fs (brown arrows in \autoref{Dynamic}).

Following these arguments, it can be understood why the FGR model, in which all vibrational modes are taken into account, works better than the TLS and SVM models at sufficiently long times, as observed in \autoref{Dynamic}. However, FGR assumes that the vibrational modes at all times are in thermal equilibrium and thus neglects the initial strongly non-equilibrium state and its coherent wave packet motion. The FGR approach thus fails to capture the short-time dynamics, and in particular cannot represent the weak oscillations observed in the TN calculations.

\begin{figure}
\includegraphics[width=\columnwidth]{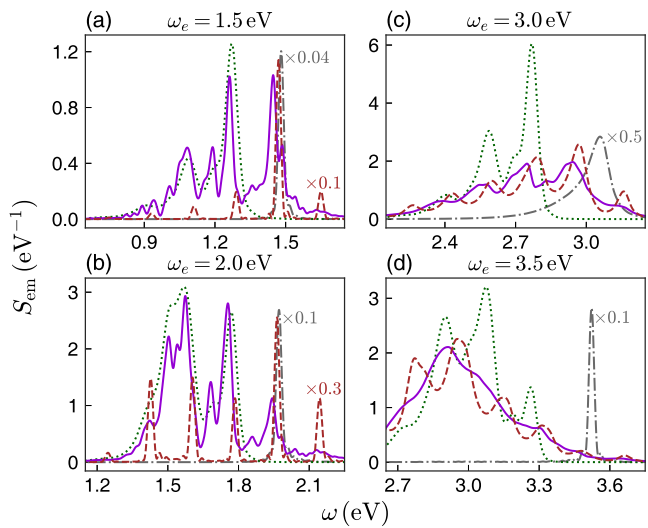}
\caption{\label{Spectra} Near-field emission spectra $S_{\mathrm{em}}(\omega)$ for four different exciton frequencies. The spectra for the TLS and SVM approximations and full TN method are calculated at $t=150$ fs. The line code is the same as in \autoref{Dynamic}. Some spectra have been scaled to facilitate comparison.}
\end{figure}

We next examine the frequency-dependent population of the EM modes, $S_{\mathrm{em}}(\omega) = \sum_{i} \langle a_{i}^{\dagger}(\omega)a_{i}^{}(\omega) \rangle$. Notice that the sum extends over both radiative and non-radiative plasmonic modes, so we name this physical magnitude the {\it near-field emission spectrum}. Within the FGR approach, this quantity is independent of time and can be written as $S_{\mathrm{em}}(\omega) \propto D_{\mathrm{em}}(\omega) J_{\mathrm{p}}(\omega)$ \cite{SM}. Figure~\ref{Spectra} shows the near-field emission spectra calculated for the four exciton frequencies and the four theoretical approaches in \autoref{Dynamic}, evaluated at $t=150$~fs. Within the TLS model, only the EM modes close to $\omega_{e}$ contribute to the spontaneous emission process. This leads to single-peaked spectra broadened by the plasmonic environment, very different to the spectra obtained within the TN framework. On the other hand, the SVM approximation largely fails to reproduce the TN-spectra for low exciton frequencies ($1.5$ and $2.0$~eV) although it provides a reasonable approximation for higher $\omega_e$. This is in accordance with its better accuracy describing the excited-state decay dynamics for those frequencies (see \autoref{Dynamic}).

In principle, one could have expected that the FGR approach should work better than the other two simplified models, as it incorporates the whole vibration spectrum via the lineshape function of the organic molecule. However, as observed in \autoref{Spectra}, this is not the case for the four chosen $\omega_e$. The FGR approach only works well when the coupling of the molecular exciton with its EM environment is weak enough such that vibrational thermal equilibrium is reached before emission takes place. However, for large plasmon-exciton couplings leading to huge Purcell factors as those associated with plasmonic fields, the molecular exciton decays so fast that vibrational equilibrium is not reached and the spectrum is significantly different from the stationary limit. In agreement with recent experiments~\cite{Sandoghdar2019, Menon2020}, our TN calculations show that this fast decay can be utilized to strongly modify the branching ratio of the emission by organic molecules.

\begin{figure}
\includegraphics[width=\columnwidth]{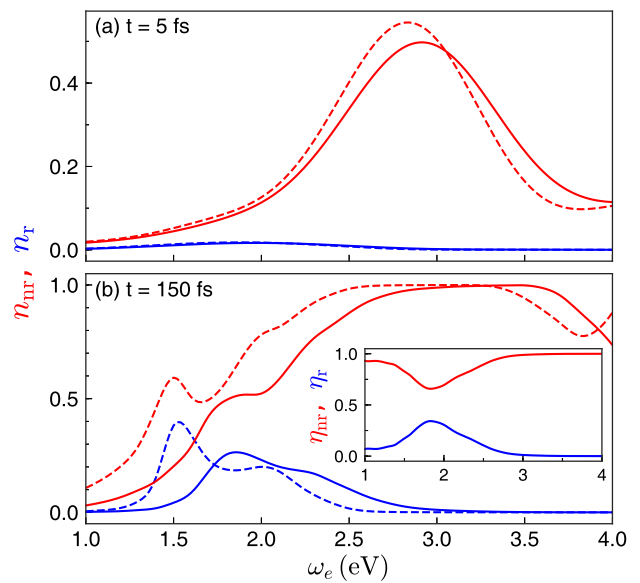}
\caption{\label{Quenching} Radiative and non-radiative plasmon population ($n_{\textrm {nr}}$ and $n_{\textrm{r}}$) as a function of exciton frequency $\omega_{e}$ at (a) t = 5 fs and (b) t = 150 fs, calculated from the TLS approximation (dashed) and full TN framework (solid). The inset in panel (b) shows the relative contributions of the radiative (effective quantum yield) and non-radiative plasmonic modes.}
\end{figure}

Finally, taking advantage of the capability of our TN theoretical framework to separate contributions from plasmonic radiative and non-radiative channels, here we address the interplay between the Purcell effect and the quenching phenomenon. Figure~\ref{Quenching} shows the evolution of the population of EM modes, $n_{i}=\int d\omega \langle a_{i}^{\dagger}(\omega)a_{i}^{}(\omega) \rangle$ ($i$ being either $\textrm{nr}$ or $\textrm{r}$), as a function of exciton frequency and evaluated at two different times. At very short time scales ($t=5$~fs, panel a), the TLS approximation reproduces the TN result. The non-radiative plasmon population is maximum at exciton frequencies at resonance with the pseudomode (around $2.9$~eV). The radiative plasmon population becomes relevant at longer times, as shown in \autoref{Quenching}(b). In the TLS approximation, this population is maximum when the exciton frequency coincides with the dipolar mode ($1.5$ eV). On the contrary, the TN calculations reveal that, the maximum far-field emission takes place when the exciton frequency is slightly blue-detuned. This setup maximizes the spectral overlap between $D_{\mathrm{em}}(\omega)$ and $J_{\mathrm{r}}(\omega)$. Within this exciton frequency range, the effective quantum yield, shown in the inset of \autoref{Quenching}(b), can be as high $30\%$, despite the fact that the gap in the plasmonic cavity is $1$ nm wide and quenching is expected to be dominant.  For higher $\omega_e$, non-radiative components take over, completely quenching the far-field emission of photons by the molecule, as expected.

To conclude, we have applied an accurate quantum tensor network method, able to treat electronic, electromagnetic and vibrational degrees of freedom in light-matter scenarios on an equal footing, to study the Purcell effect occurring when organic molecules interact with plasmonic nanocavities. Using this numerical framework we have tested different simplified models that have been used to analyse this phenomenon lately. We have found that a two-level description of the molecule fairly captures the first steps of the decay process and a simple approach based on the Fermi golden rule can describe its very long time behavior. However, an accurate treatment of the exciton coupling to all the vibrational modes in the organic molecule is mandatory to resolve intermediate timescales. As a consequence, none of the simplified models used up to date are able to provide a precise estimation of near- or far-field emission spectra in hybrid systems involving organic molecules and plasmonic cavities. As an interesting extension of the current work, the influence of non-linear vibrational mode couplings leading to vibrational energy redistribution could be investigated, as they are not included in the current Holstein-type model and are expected to become important on slightly longer timescales than treated here \cite{Clear2020,Reitz2019}. Our findings also reveal that significant effective quantum yield values can be achieved in situations where strong interaction to non-radiative plasmonic modes takes place and quenching is expected to dominate. 

\begin{acknowledgments}
This work has been funded by the National Natural Science Foundation of China
under Grant No.~11804283, by the European Research Council through grant
ERC-2016-STG-714870, and by the Spanish Ministry for Science and Innovation
-- AEI grants RTI2018-099737-B-I00, PCI2018-093145 (through the
QuantERA program of the European Commission), and CEX2018-000805-M (through the ``Maria de Maeztu'' Programme for Units of Excellence in R\&D).
AIFD acknowledges support from a 2019 Leonardo Grant for Researchers and Cultural Creators, BBVA Foundation. D.~Zhao acknowledges
financial support from the China Scholarship Council to fund his stay at
Universidad Aut\'{o}noma de Madrid as a postdoctoral fellow.
\end{acknowledgments}

\bibliography{Ref}

\end{document}


\title{Supplemental Material: \\Plasmonic Purcell Effect in Organic Molecules}
\author{D. Zhao$^{1,2}$}
\author{R. E. F. Silva$^2$}
\author{C. Climent$^2$}
\author{J. Feist$^2$}
\author{A. I. Fern\'{a}ndez-Dom\'{\i}nguez$^2$}
\author{F. J. Garc\'{\i}a-Vidal$^{2,3}$}
\affiliation{
$^1$School of Physical Science and Technology, Southwest University, Chongqing 400715, China \\
$^2$Departamento de F\'{i}sica Te\'{o}rica de la Materia Condensada and Condensed Matter Physics Center (IFIMAC), Universidad Aut\'{o}noma de Madrid, E-28049 Madrid, Spain\\
$^3$Donostia International Physics Center (DIPC), E-20018 Donostia/San Sebasti\'{a}n, Spain}


\date{\today}
%

\maketitle

\section{tensor network calculations}
Given that the plasmonic environment consists of both radiative and non-radiative parts, one may guess that the decay dynamics depend on the respective values of each components. However, as shown in the following, the excited-state dynamics of the molecules are governed by the total EM spectral density $J_{\mathrm{p}}(\omega)$ only. To this end, starting from Eq. (1) in the main text, we first introduce two sets of new plasmonic modes
\begin{eqnarray}
a_{\omega} &=& \alpha_{\mathrm{nr},\omega} a_{\mathrm{nr}, \omega} + \alpha_{\mathrm{r},\omega} a_{\mathrm{r}, \omega}, \\
d_{\omega} &=& \alpha_{\mathrm{r},\omega} a_{\mathrm{nr}, \omega} - \alpha_{\mathrm{nr},\omega} a_{\mathrm{r}, \omega},
\end{eqnarray}
where the coefficients $\alpha_{i,\omega}$ are defined by $\alpha_{i,\omega} = g_{i,\omega}/g_{\omega}$, with $g_{\omega} = \sqrt{g_{\mathrm{nr},\omega}^{2}+g_{\mathrm{r},\omega}^{2}}$. Rewriting Eq. (1) in the main text with the new operators gives
\begin{eqnarray}
H &=&   \omega_{e}\sigma_{+}\sigma_{-} \nonumber + \int d\omega \left [\omega b^{\dagger}_{\omega} b^{}_\omega + \lambda_{\omega}\left (b^{\dagger}_{\omega} +  b^{}_{\omega}\right )\sigma_{+}\sigma_{-} \right] \nonumber \nonumber\\
& &  + \int d\omega \left[\omega a_{\omega}^{\dagger}a_{\omega}^{} + \omega d_{\omega}^{\dagger}d_{\omega}^{} + g_{\omega} \left (a_{\omega}^{\dagger}\sigma_{-}+a_{\omega}^{}\sigma_{+}\right) \right].\nonumber\\
\end{eqnarray}
Here one can see that $d_{\omega}$ acts as a \textit{dark} mode that is decoupled from the exciton. Moreover, the original EM modes can be expressed in terms of the new mode operators as
\begin{eqnarray}
a_{\mathrm{nr},\omega} &=& \alpha_{\mathrm{nr},\omega} a_{\omega} + \alpha_{\mathrm{r},\omega} d_{\omega}, \\
a_{\mathrm{r},\omega} &=& \alpha_{\mathrm{r},\omega} a_{\omega} - \alpha_{\mathrm{r},\omega} d_{\omega}.
\end{eqnarray}
Taking the non-radiative component as an example, the population in this branch can be calculated from
\begin{eqnarray}
\langle a_{\mathrm{nr},\omega}^{\dagger} a_{\mathrm{nr},\omega}^{}\rangle &=& \alpha_{\mathrm{nr},\omega}^{2}\langle a_{\omega}^{\dagger}a_{\omega}^{}\rangle + \alpha_{\mathrm{r},\omega}^{2}\langle d_{\omega}^{\dagger}d_{\omega}^{}\rangle \nonumber \\
& & + \alpha_{\mathrm{nr},\omega}\alpha_{\mathrm{r},\omega}(\langle a_{\omega}^{\dagger}d_{\omega}^{}\rangle + \langle a_{\omega}^{}d_{\omega}^{\dagger}\rangle).
\end{eqnarray}
Since we assume an initial vacuum state for the EM modes and mode $d_{\omega}$ is decoupled from the exciton, any expectation value related with $d_{\omega}$ vanishes. With this argument we arrive to the following expression for both the radiative and non-radiative components, 
\begin{eqnarray}
\langle a_{i,\omega}^{\dagger}a_{i,\omega}^{}\rangle = \alpha_{i,\omega}^{2}\langle a_{\omega}^{\dagger}a_{\omega}^{}\rangle = \frac{J_{i}(\omega)}{J_{\mathrm{p}}(\omega)}\langle a_{\omega}^{\dagger}a_{\omega}^{}\rangle,
\end{eqnarray} 
with the spectral density $J_{i}(\omega)=g_{i,\omega}^{2}$ and $J_{\mathrm{\mathrm{p}}}(\omega)=g_{\omega}^{2}=J_{\mathrm{nr}}(\omega) + J_{\mathrm{r}}(\omega)$.
Therefore, the excited-state dynamics of the molecules depends only on the $a_{\omega}$ operators with coupling strength $g_{\omega}$, and hence the total EM spectral density $J_{\mathrm{\mathrm{p}}}(\omega)$.

To perform the tensor network (TN) calculations, we first rewrite the system Hamiltonian into a discrete form
\begin{eqnarray}
\label{DHamiltonian}
H_{\mathrm{d}} &=&  \omega_{e}\sigma_{+}\sigma_{-} \nonumber + \sum_{k=1}^{K} \left [\omega_{k}^{\mathrm{v}} b^{\dagger}_{k} b_{k}^{} + \lambda_{k}^{}\left (b^{\dagger}_{k} +  b^{}_{k}\right )\sigma_{+}\sigma_{-} \right] \nonumber\\
& & + \sum_{l=1}^{L} \left[\omega_{l}^{\mathrm{p}} a_{l}^{\dagger}a_{l}^{} + g_{l}^{} \left (a_{l}^{\dagger}\sigma_{-}+a_{l}^{}\sigma_{+}\right) \right],
\end{eqnarray}
where we have omitted the $d_{\omega}$ operators for simplicity. Note that after the discretization, $b_{k}$ and $a_{l}$ are dimensionless operators, while $\lambda_{k}$ and  $g_{l}$ have the same dimension as frequency. The spectra $S_{\mathrm{em}}(\omega)=\langle a_{\omega}^{\dagger}a_{\omega}^{}\rangle$ defined in the main text can be obtained from  $S_{\mathrm{em}}(\omega)=\langle a_{l}^{\dagger}a_{l}^{}\rangle / \Delta \omega$, with $\Delta \omega$ being the frequency interval for the discretization. 

To enable the treatment of Eq. (\ref{DHamiltonian}) by the density matrix renormalization group methods \cite{Ulrich2011}, a crucial step is applying a chain mapping transformation. It transforms the Hamiltonian into a form containing one-dimensional chains with only nearest-neighbor interactions \cite{Alex2010, Alex2011}. By doing so the system Hamiltonian can be written in an equivalent chain form
\begin{eqnarray}
\label{CHamiltonian}
&H_{\mathrm{c}}&=  \omega_{e}\sigma_{+}\sigma_{-} \nonumber \\
&+& \sum_{k=1}^{K}\tilde{\omega}_{k}^{\mathrm{v}}\tilde{b}_{k}^{\dagger}\tilde{b}_{k}^{}+ t^{\mathrm{v}}_{1}(\tilde{b}_{1}^{\dagger}+\tilde{b}_{1}^{})\sigma_{+}\sigma_{-} + \sum_{k=2}^{K}t_{k}^{\mathrm{v}}(\tilde{b}_{k-1}^{\dagger}\tilde{b}_{k}^{}+\mathrm{H.c.}) \nonumber\\
&+& \sum_{l=1}^{L}\tilde{\omega}_{l}^{\mathrm{p}}\tilde{a}_{l}^{\dagger}\tilde{a}_{l}^{}+ t^{\mathrm{p}}_{1}(\tilde{a}_{1}^{\dagger}\sigma_{-}+\tilde{a}_{1}^{}\sigma_{+}) + \sum_{l=2}^{L}t_{l}^{\mathrm{p}}(\tilde{a}_{l-1}^{\dagger}\tilde{a}_{l}^{}+\mathrm{H.c.})\nonumber. \\
\end{eqnarray}
Therefore, the exciton is now coupled to the vibrational and plasmonic chains. The former accounts for the dressing of the exciton by the molecular vibrations, while the latter leads to the exciton decay. Among many other chain parameters, $\tilde{\omega}_{1}^{\mathrm{v}}$, $t_{1}^{\mathrm{v}}$ and $t_{2}^{\mathrm{v}}$, denoted by $\omega_{\mathrm{RC}}$,  $\lambda_{\mathrm{RC}}$ and $\gamma_{\mathrm{d}}$ respectively in the main text, are of vital importance for determining the timescales where the TLS and SVM approximations work. 

The Hamiltonian in Eq. (\ref{CHamiltonian}) is formidable with a brute-force approach for a large number of modes. By expressing the system's wave function as a TN, i.e., a network of interconnected tensors with fewer coefficients, the computation demand can be greatly reduced \cite{Orus2019}. Therefore, TN methods provides an efficient way to simulate many-body dynamics and allow for the computation of a quasi-exact solution to the time-dependent Schr\"{o}dinger equation for very complex systems. In this work we calculate the dynamics with the time-dependent variational matrix product states (TDVMPS) technique \cite{Haegeman2016, DelPino2018PRB, Florian2019}. 

The main parameters for the TN calculations in this work are as follows. For the plasmonic part, the EM modes with frequency range from 0.6 eV to 6.0 eV have been considered, while for the molecular vibrations, all the modes of the CPDT dye have been included, lying within 0.1 to 500 meV. Both plasmonic and vibrational environment are discretized into $L=M=650$ modes. Moreover, to perform the TN calculations, a bond dimension $D = 22$ and a time step $\Delta t = 5$ a.u. have been used. All these parameters have been checked to guarantee the convergence of the results. 

With regard to the organic molecule, electronic structure calculations within density functional theory and its time-dependent version were performed to obtain the vibronic coupling constants for the displaced harmonic oscillator model. The long-range corrected CAM-B3LYP functional \cite{Yanai2004} was used together with the 6-31G(d) basis set to properly describe the charge-transfer transition. For the most stable molecular conformer of the CPDT dye \cite{Climent2013}, ground and excited state harmonic frequencies were calculated and the Huang-Rhys factors were obtained considering Duschinsky rotation effects \cite{Duschinsky1937}. All calculations were performed with the Gaussian 16 package \cite{Gaussian2016}. Note that the CPDT dye is closely related to the commercial C218, Y123 and Dynamo Red dyes widely employed in dye sensitized solar cells.

\section{Fermi\rq{}s golden rule}
According to the Fermi\rq{}s golden rule (FGR), if a quantum emitter [transition frequency $\omega_{0}$] is weakly coupled to a set of modes ${\alpha}$ [mode frequency $\omega_{\alpha}$] with a coupling strength $g_{\alpha}$, then the irreversible decay rate of the emitter can then be calculated from 
\begin{equation}
 \gamma_{0} = 2\pi \sum_{\alpha} g_{\alpha}^{2}\delta(\omega_{0}-\omega_{\alpha}).
\end{equation}
In the following, we apply this rule to the excited-state dynamics of the organic molecules. 

Assuming that, initially, the molecule is in its electronic excited state and all plasmonic modes are in their vacuum state, one can work in the single-excitation subspace for the spontaneous emission process. Using the completeness of the vibrational basis, the Hamiltonian in  Eq. (\ref{DHamiltonian}) can be rewritten as
\begin{eqnarray}
H_{\mathrm{d}} &=& \sum_{M} \omega_{e,M}|e, 0, M\rangle \langle e, 0, M| \nonumber\\ 
 & & + \sum_{l} \sum_{N} \omega_{g,l,N}|g, 1_{l}, N\rangle \langle g, 1_{l}, n|  \\
 & & + \sum_{l}\sum_{M,N} g_{l} \langle N| M\rangle (|e, 0, M\rangle \langle g, 1_{l}, N| + \mathrm{H.c.}), \nonumber
\end{eqnarray}
with $M$ ($N$) representing a set of quantum numbers $\{m_{1}, m_{2}, \ldots, m_{k}, \ldots\}$ ($\{n_{1}, n_{2}, \ldots, n_{k}, \ldots\}$) for the vibrational basis belonging to the electronic \textit{excited} (\textit{ground}) state. The basis $|e, 0\rangle$ ($|g, 1_{l}\rangle$) denotes the electronic excited (ground) state, and vacuum state for all the cavity modes (one-photon state for cavity mode $l$ and vacuum state for the remaining). Moreover, the corresponding energy for state $|e, 0, M\rangle$ and $|g, 1_{l}, N\rangle$ is $\omega_{e,M}=\omega_{e}-\Delta_{\mathrm{re}}+\sum_{k}m_{k}\omega_{k}^{\mathrm{v}}$ and $\omega_{g,l,N}=\omega_{l}^{\mathrm{p}}+\sum_{k}n_{k}\omega_{k}^{\mathrm{v}}$, respectively. The the reorganization energy can be calculated as $\Delta_{\mathrm{re}}=\sum_{k}\lambda_{k}^{2}/\omega_{k}^{\mathrm{v}}$ with the discrete parameters. The spontaneous emission process is manifested through the coupling of the manifold of initial states $\{|e, 0, M\rangle\}$ to the manifold of final states $\{|g, 1_{l}, N\rangle\}$ with strength $g_{l} \langle N| M\rangle$. The factor $\langle N| M\rangle = \Pi_{k}\langle n_{k}|m_{k}\rangle$, where $\langle n_{k}|m_{k}\rangle$, known as Franck-Condon factor, is the overlap integral between the vibrational wave functions $|n_{k}\rangle$ and $|m_{k}\rangle$.

According to the FGR, if the molecule is initially in the state $|e,0,M\rangle$, then the decay rate from this state to the manifold of final states $\{|g, 1_{l}, N\rangle\}$ reads
\begin{equation}
\gamma_{M} = 2\pi \sum_{l}\sum_{N} g^{2}_{l}\langle M|N\rangle^{2} \delta(\omega_{e,M}-\omega_{g,l,N}).
\end{equation}
The above equation can be written as
\begin{equation}
\label{gamma_m}
\gamma_{M} = 2 \pi \int d\omega D_{M}(\omega) J_{\mathrm{p}}(\omega)
\end{equation}
after introducing a lineshape function $D_{M}(\omega)$ with
\begin{equation}
\label{D_m}
D_{M}(\omega) = \sum_{N} \langle M|N\rangle^{2} \delta\left(\omega_{e,M} - \omega_{g,l,N} +\omega_{l}^{\mathrm{p}}-\omega\right)
\end{equation}
for a specific initial state $|e,0,M\rangle$. The corresponding \textit{near-field emission spectra} introduced in the main text can then be calculated as
\begin{equation}
S_{M}(\omega) = \frac{2\pi}{\gamma_{M}} (1-e^{-\gamma_{M}t})D_{M}(\omega)J_{\mathrm{p}}(\omega),
\end{equation}
where integrating over $\omega$ yields the population in the ground state. However, the initial state of the molecule is usually a mixture of states in the manifold $\{|e, 0, M\rangle\}$, with the probability of state $|e, 0, M\rangle$ being $f_{e,M}$ which fulfils $\sum_{M}f_{e,M}=1$. In this situation, two limiting cases are usually considered: 

(\textit{i}) The decay processes starting from each of the states in $\{|e, 0, M\rangle\}$ are independent, which leads to a multi-exponential decay dynamics with $\langle \sigma_{+}\sigma_{-}\rangle =\sum_{M}f_{e,M} e^{-\gamma_{M}t}$. In this case, the near-field emission spectrum$S_{\mathrm{em}}^{(i)}(\omega)$ can be obtained from
\begin{equation}
S_{\mathrm{em}}^{(i)}(\omega) = \sum_{M}f_{e,M}S_{M}(\omega);
\end{equation}

(\textit{ii}) Fast phonon redistribution takes place, by thermalization for example, with a typical timescale much faster than $1/\gamma_{M}$. This leads to a fixed probability distribution among states $\{|e, 0, M\rangle\}$ during the whole decay process. In this case, the decay of the excited-state population is given by a single exponential as $\langle \sigma_{+}\sigma_{-}\rangle = e^{-\gamma t}$, with the decay rate being
\begin{eqnarray}
 \gamma = \sum_{M} f_{e,M}\gamma_{M}.
\end{eqnarray}
The above expression can also be rewritten as
\begin{equation}
\label{gamma}
\gamma = 2 \pi \int d\omega D_{\mathrm{em}}(\omega) J_{\mathrm{p}}(\omega),
\end{equation}
with the averaged lineshape function $D_{\mathrm{em}}(\omega)$ given by
\begin{equation}
\label{D_em}
D_{\mathrm{em}}(\omega) = \sum_{M} f_{e,M}D_{M}(\omega).
\end{equation}
In this case, the near-field emission spectrum reads
\begin{equation}
S_{\mathrm{em}}^{(ii)}(\omega) = \frac{2\pi}{\gamma}(1-e^{-\gamma t}) D_{\mathrm{em}}(\omega) J_{\mathrm{p}}(\omega).
\end{equation}
Strictly speaking, as we don\rq{}t include any phonon redistribution mechanism in our Hamiltonian Eq. (\ref{DHamiltonian}), the decay should follow the multi-exponential dynamics discussed in case (\textit{i}). However, as we will show below, there is no significant difference between both cases for the studied system.

\begin{figure}
	\center
	\includegraphics[width=8.6cm]{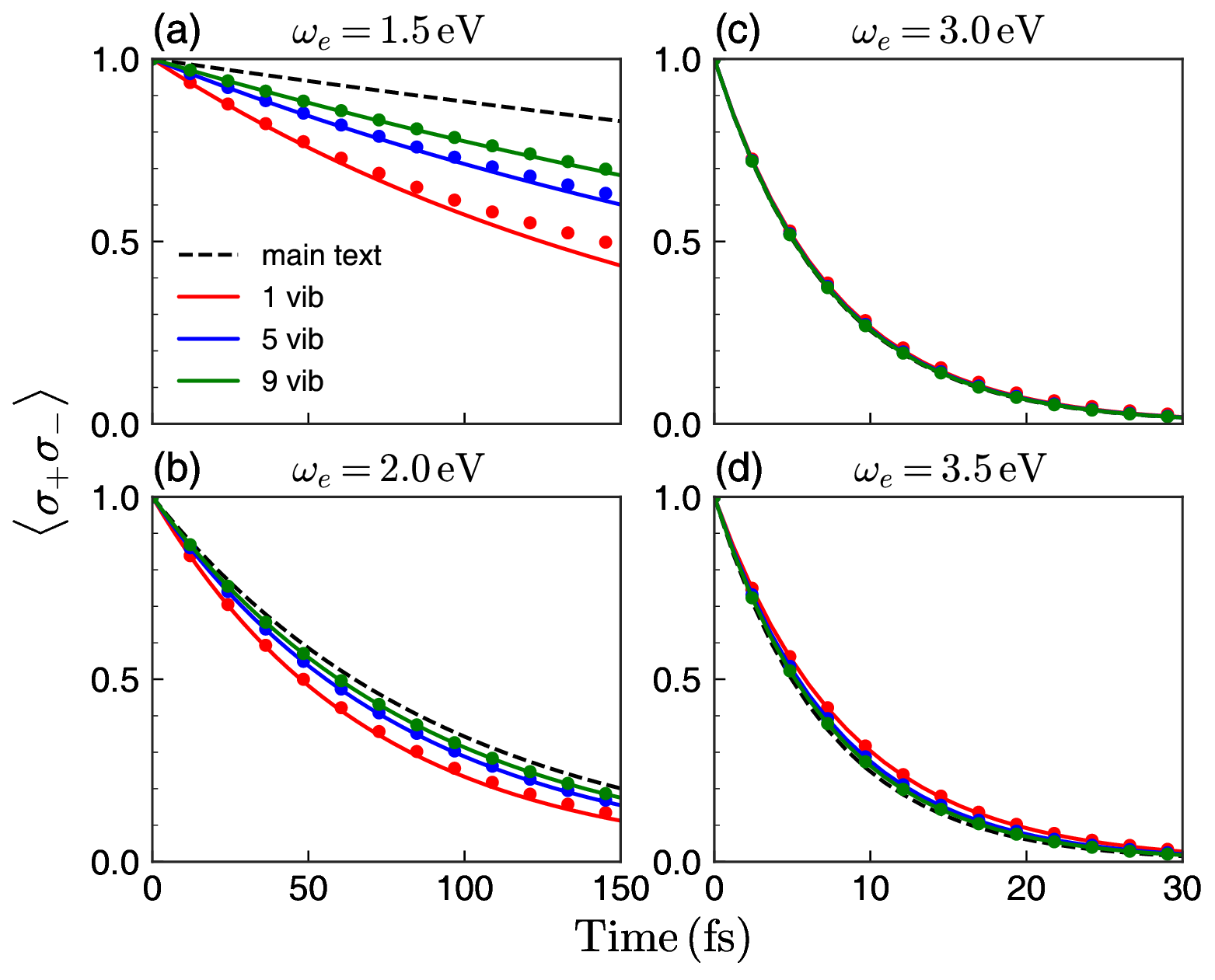}
	\caption{\label{Fig_S1} The excited-state dynamics $\langle \sigma_{+} \sigma_{-} \rangle$ calculated from multi-exponential [case $(\textit{i})$, dot scatters] and single-exponential dynamics [case $(\textit{ii})$, solid curves] with one (red), five (blue), and nine (green) vibrational modes taken into account for the exciton frequencies discussed in the main text. The parameters for vibrational modes are obtained from an inverse chain mapping method by keeping only the first few chain sites. The FGR results considering all the vibrational modes in the main text are also shown (dashed curves).}
\end{figure}

To calculate the decay rate in Eq. (\ref{gamma_m}) and (\ref{gamma}) and hence the lineshape function, detailed knowledge of the vibrational eigenstates and eigenenergies for both the excited and ground state potential energy surfaces is required. Here we focus on the case that the molecular vibrations can be approximately described by displaced harmonic oscillators as in Eq. (\ref{DHamiltonian}). With respect to the initial state, we consider a Franck-Condon excitation, the vertical transition from the vibrational ground state of the electronic ground state by ultrashort laser pulses. This leads to an initial distribution
\begin{equation}
\label{f_em_FC}
f_{e,M}=\prod_{k} \frac{1}{m_{k}!}\left(\frac{\lambda_{k}}{\omega_{k}^{\mathrm{v}}}\right)^{2m_{k}}\exp\left[-\left(\frac{\lambda_{k}}{\omega_{k}^{\mathrm{v}}}\right)^{2}\right].
\end{equation} 

Now we compare the two limiting cases mentioned above. Figure \ref{Fig_S1} shows the comparison between the multi-exponential and single-exponential dynamics for an initial Franck-Condon excitation. To ease the numerical computation, only a few vibrational modes have been included. The two cases show slight differences if we consider one vibrational mode only. By increasing the number of vibrational modes, the calculations tend to converge and the differences between those two cases disappear. Moreover, due to the broadband structure and strong Purcell enhancement effect of the plasmonic pseudomode, it is easier to converge calculations for excitons coupled to this mode, meaning that the decay rates are not sensitive to the lineshape function. In contrast, the relatively narrower structure of the lower-order modes necessitate a more accurate description of the lineshape function, and hence the inclusion of more vibrational modes.

\begin{figure*}
	\center
	\includegraphics[width=16cm]{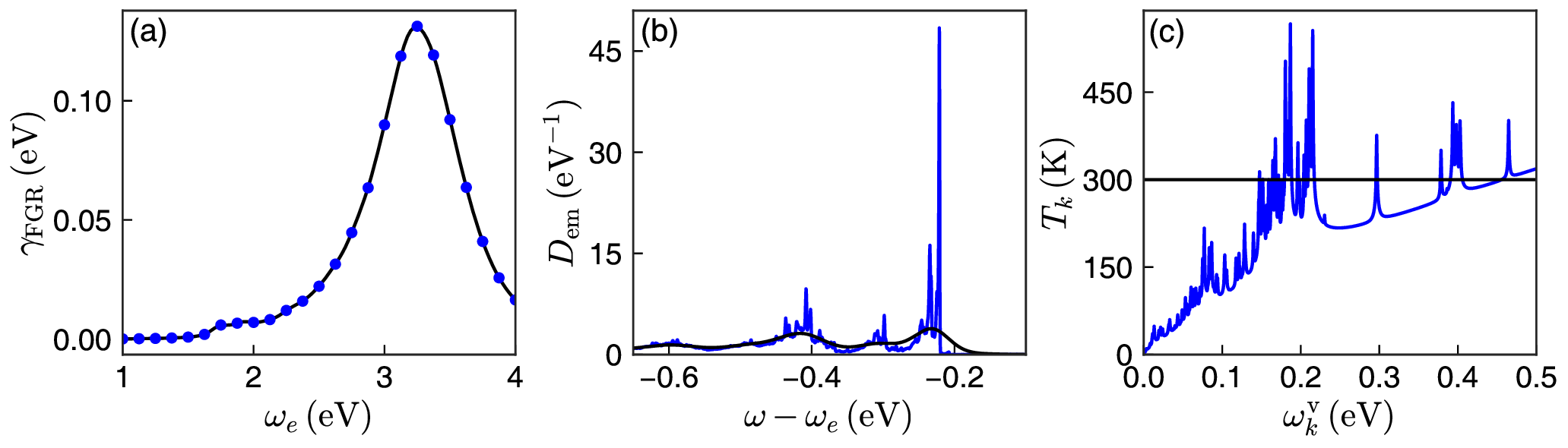}
	\caption{\label{Fig_S2} (a) Decay rates and (b) lineshape function calculated with the assumption of $T=300$ K (black) and with pseudo temperatures (blue). (c) Pseudo temperatures $T_{k}$ (blue) used to simulate the Franck-Condon distribution with a thermal distribution. The black horizontal line shows the temperature used for the FGR results in the main text.}
\end{figure*}

Although the brute-force calculation of the lineshape function based on Eq. (\ref{D_m}) and (\ref{D_em}) works quite well for a few number of vibrational modes, they are prohibitive for hundreds of modes, which is the typical number of vibrational modes for organic molecules. Fortunately, if fast thermalization takes place for the molecular vibrations due to its interaction with a thermal reservoir at temperature $T$, an analytical expression for the lineshape function can be derived as \cite{May2011}
\begin{equation}
\label{D_Fourier}
D_{\mathrm{em}}(\omega) = \frac{1}{2\pi} \int dt e^{-i[\omega-(\omega_{e}-\Delta_{\mathrm{re}})]t-G(0)+G(t)}.
\end{equation}
The time-dependent function in the exponent reads
\begin{equation}
G(t) = \sum_{k} \left(\frac{\lambda_{k}}{\omega_{k}^{\mathrm{v}}}\right)^{2}\{\left[1+n(\omega_{k}^{\mathrm{v}})\right]e^{-i\omega_{k}^{\mathrm{v}}t} + n(\omega_{k}^{\mathrm{v}})e^{i\omega_{k}^{\mathrm{v}}t}\},
\end{equation}
with $n(\omega_{k}^{\mathrm{v}})=\left[\exp(\hbar\omega_{k}^{\mathrm{v}}/k_{\mathrm{B}}T)-1\right]^{-1}$ being the Bose–Einstein distribution function assuming a thermalized initial distribution 
\begin{equation}
\label{f_em_T}
f_{e,M}=\prod_{k}\exp\left(-\frac{m_{k}\hbar\omega_{k}^{\mathrm{v}}}{k_{\mathrm{B}}T}\right)\left[1-\exp(-\frac{\hbar\omega_{k}^{\mathrm{v}}}{k_{\mathrm{B}}T})\right],
\end{equation}
in contrast to the Franck-Condon distribution. Interestingly, it is possible to introduce a pseudo temperature $T_{k}$ for each vibrational mode $k$ to simulate the Franck-Condon distribution with a thermal distribution, as long as the factor $\lambda_{k}/\omega_{k}^{\mathrm{v}}$ is small, say, less than 0.3. This pseudo temperature is found to be 
\begin{equation}
\label{PseudoT}
k_{\mathrm{B}}T_{k}=-\frac{\hbar\omega_{k}^{\mathrm{v}}}{\ln \{1-\exp[-\left(\lambda_{k}/\omega_{k}^{\mathrm{v}}\right)^{2}]\}},
\end{equation}
which is obtained by forcing the thermal distribution to have the same vibrational ground state population as the Franck-Condon distribution.

In the main text, the lineshape function and the FGR results are based on Eqs. (\ref{D_Fourier})-(\ref{f_em_T}) with the assumption of fast thermalization at $T=300$ K. To justify this choice, we compare the decay rates as well as the lineshape function calculated at $T=300$ K and pseudo temperature $T_{k}$ in Fig. \ref{Fig_S2}. Excellent agreement for the decay rates can be found. As for the lineshape function, except for the sharp vibronic structures shown in the pseudo temperature case, the main peak positions are also in good agreement.

\bibliography{Ref}